\documentclass{ametsocV6.1}

\title{Marine Heatwaves in the Arabian Sea: Drivers and Impacts on Atmospheric Circulation and Extreme Precipitation}

\authors{D. L. Suhas\aff{1*}, Weiqing Han\aff{1}, Toshiaki Shinoda\aff{2}, Rui Sun\aff{3}, Aneesh Subramanian\aff{1}, Mark Bourassa\aff{4,5}, Michael Alexander\aff{1} \correspondingauthor{D. L. Suhas, suhasdl@colorado.edu \\ \textit{Under review at \textbf{Journal of Climate}}}
}

\affiliation{
\aff{1}{Department of Atmospheric and Oceanic Sciences, University of Colorado, Boulder, CO, USA}\\
\aff{2}{Department of Physical and Environmental Sciences, Texas A\&M University--Corpus Christi, Corpus Christi, TX, United States}\\
\aff{3}{Scripps Institution of Oceanography, University of California, San Diego, La Jolla, CA, USA}\\
\aff{4}{Department of Earth, Ocean and Atmospheric Science, Florida State University, Tallahassee, FL, USA}\\
\aff{5}{Center for Ocean-Atmospheric Prediction Studies (COAPS), Florida State University, Tallahassee, FL, USA}\\
}

\abstract{
Marine heatwaves (MHWs) threaten marine ecosystems and significantly impact weather patterns. In the Arabian Sea, summer MHWs are of particular concern due to their potential impacts on the Indian summer monsoon, a lifeline for nearly a billion people. However, the drivers of these MHWs and their influence on atmospheric circulation and monsoon rainfall remain poorly understood. Using satellite observations, reanalysis datasets, and numerical model experiments, we investigate the key drivers of MHW events and assess their impacts. When SST warming trends are retained, the eastern and northern Arabian Sea emerge as MHW hotspots, showing rapid increases during 1982--2023, largely due to anthropogenic warming. On detrending the SSTs to remove the influence of anthropogenic warming on individual MHWs, we find that most MHWs are short-lived (lasting $\le$ 20 days) and are initiated by enhanced surface shortwave radiation and reduced latent heat loss associated with the suppressed convection phase of the Boreal Summer Intraseasonal Oscillations (BSISOs). Interannual SST anomalies, including ENSO and Indian Ocean Dipole (IOD), further modulate the year-to-year MHW variability. Conversely, the warm SSTs during MHWs exert strong atmospheric feedbacks. MHWs in the eastern Arabian Sea drive cyclonic winds, intensify moisture convergence and increase the risk of extreme precipitation along the southwest coast of India. In the northern Arabian Sea, MHW-induced cyclones trigger intense rainfall over northwestern India and Pakistan, contributing to extreme events like the 2022 Pakistan floods. These findings improve our capacity to predict Arabian Sea MHWs and assess their risks, offering significant socio-economic and ecological benefits.
}

\begin{document}

\maketitle

\statement
This study investigates the causes of marine heatwaves (MHWs) in the Arabian Sea during the summer monsoon season and their role in enhancing extreme rainfall over large parts of India and Pakistan. This is the first study to link MHWs with extreme rainfall in the region, including events such as the 2022 Pakistan floods. It highlights the two-way interaction between the atmosphere and ocean---where atmospheric conditions drive MHWs, and MHWs in turn influence atmospheric circulation and precipitation. These findings advance our understanding of Arabian Sea MHWs and support efforts to predict extreme events and assess associated risks, with important socio-economic and ecological implications.

\section{Introduction}

Marine heatwaves (MHWs) in various regions of the global oceans have been extensively studied in recent years due to their profound socio-economic and ecological impacts \citep{frolicher2018emerging,oliver2019projected,smith2021socioeconomic,capotondi2024global}. These extreme ocean warming events can last from a few days to several months, causing devastating effects on marine ecosystems \citep{smale2019marine,wernberg2013extreme,cavole2016biological,smith2023biological}, including widespread coral bleaching \citep{le2017marine,fordyce2019marine}, range shifts and mass mortality of marine species \citep{smale2013extreme,wernberg2016climate,garrabou2009mass}, and disruptions to fisheries with significant consequences for coastal communities \citep{mills2013fisheries,cheung2020marine,smith2021socioeconomic}. While MHWs are driven by atmospheric variability combined with oceanic factors \citep{vogt2022local,bian2024scale}, they can also influence the atmosphere, impacting tropical cyclones  \citep{rathore2022interactions,choi2024marine,radfar2024rapid} and regional precipitation \citep{saranya2022genesis}. In the presence of the long-term global warming trend, regions experiencing unprecedented warming in recent decades witness more frequent and intense MHWs \citep{oliver2018longer,frolicher2018marine,laufkotter2020high, chatterjee2022marine, saranya2022genesis, xu2022increase}, particularly when a fixed sea surface temperature (SST) threshold is used to identify MHW events \citep{amaya2023marine,smith2025baseline}.

In the Indian Ocean, recent studies have examined MHWs in a limited number of regions, particularly the Bay of Bengal \citep{saranya2022genesis, mandal2024influence}, the western equatorial basin off the Somalia coast \citep{saranya2022genesis, qi2022characteristics}, and coastal areas of the Southeast Indian Ocean \citep{feng2013nina,benthuysen2018extreme,han2022sea}. These studies show that the rapid warming of SSTs has contributed to increased MHW frequency in these areas. The El Ni\~{n}o-Southern Oscillation (ENSO) as well as the Indian Ocean Dipole (IOD)---the dominant modes of interannual climate variability in the tropical Indo-Pacific basin \citep{klein1999remote,saji1999dipole, meyers2007years, chowdary2007basin}---can also modulate MHW activities \citep{saranya2022genesis, qi2022characteristics, benthuysen2018extreme, han2022sea}. Recently, atmospheric boreal summer intraseasonal oscillations \citep[BSISOs;][]{sengupta2001coherent,wang2018intraseasonal,kikuchi2021boreal} have been shown to affect MHWs in the northern Bay of Bengal \citep{mandal2024influence}. BSISOs, a prominent mode of intraseasonal variability in the tropical troposphere during boreal summer, typically form near the equator and propagate northward or northeastward. They are characterized by eight phases, each representing distinct regions of convective activity \citep{kikuchi2021boreal}. The associated wind and convection anomalies drive intraseasonal SST variations, with certain phases favoring SST warming and consequently the occurrence of MHWs. As the Arabian Sea is also strongly influenced by BSISOs during the June--September monsoon season \citep{roxy2007role,li2016intraseasonal}, these oscillations may play an important role in modulating MHWs there.

In the Arabian Sea, research focusing on MHWs remains limited. A recent study suggests that the warming trend of Arabian Sea SST has led to an increase in both the frequency and duration of MHWs over the past decades. After removing the SST trend, MHW days showed a significant correlation with ENSO but only a weak correlation with IOD \citep{chatterjee2022marine}. However, the drivers of individual MHW events are poorly understood, and the impacts of Arabian Sea MHWs on Indian summer monsoon rainfall and extreme precipitation events remain largely unknown---despite Arabian Sea SST being known to influence monsoon precipitation over India and Pakistan \citep{rao1988interannual,levine2012dependence,roman2020role, doi2024seasonal,adeel2024impact}. Yet, such understanding is essential for successfully predicting MHW events and reliably assessing the risks of extreme precipitation and flooding associated with the monsoon---a lifeline for nearly a billion people in the Indian subcontinent \citep{gadgil2003indian}---and can help facilitate more informed decision-making.

Here, we combine observational analyses and numerical model experiments to first identify the drivers of MHW events in the Arabian Sea and then assess their impacts on monsoon rainfall and extreme precipitation across regions of the Indian subcontinent. The rest of the paper is organized as follows. Section 2 describes data and methodology. In Section 3, we identify the hotspots of MHWs in the Arabian Sea---regions with a rapid increase in MHWs---and examine the role played by the warming trend of SST in these regions. In Section 4, we remove SST warming trends from each grid point before identifying MHW events to isolate the influence of internal climate variability and examine the drivers of individual events. Section 5 assesses the impacts of MHWs on atmospheric circulation and precipitation, and Section 6 provides the summary and conclusions.

\section{Data and Methods}

\subsection*{Datasets}

MHWs are identified using daily SST from the Optimum Interpolated Sea Surface Temperature (OISST) dataset, which is available at a grid spacing of $0.25^\circ \times 0.25^\circ$ for a 42-year period from 1982 to 2023 \citep{reynolds2007daily}. We focus specifically on the Indian summer monsoon season, which typically lasts from June--September \citep{gadgil2003indian}, a period when MHWs in the Arabian Sea are likely to have a direct impact on monsoon rainfall and extreme precipitation events. To assess the effects of external forcings on the SST warming trends, we analyzed outputs from 24 high-resolution CMIP6 models \citep{haarsma2016high}, including 10 ensemble members from the high-resolution CESM \citep{chang2020unprecedented}. We include all high-resolution CMIP6 models available at \url{https://aims2.llnl.gov/search/cmip6/}, considering only those with an ocean component having a horizontal grid spacing of at most $0.25^\circ$. For the period up to 2014, the historical runs (hist-1950) were used, while projections after 2014 were based on the highres-future scenario runs. The list of all high-resolution CMIP6 models used in this study is provided in Table S1 of the Supplementary Information. However, we have not used CMIP6 models to evaluate MHWs, as many of them lack the daily data needed to identify MHWs. To investigate the atmospheric drivers and impacts of MHWs, we analyze air-sea fluxes, outgoing longwave radiation (OLR), and low-level winds using ERA5 reanalysis data. Daily ERA5 data is available at a horizontal grid spacing of $0.25^\circ$ \citep{hersbach2020era5}. For comparison, we also use surface winds from the Cross-Calibrated Multi-Platform (CCMP v3.1) dataset, produced by remote sensing systems. We computed daily averages from the available 6-hourly data, which has a grid spacing of $0.25^\circ$ and is available from 1993 onward \citep{atlas2011cross}. Daily mean ECCO reanalysis data is available from 1992 to 2017, with a horizontal grid spacing of $1^\circ$ and vertical grid spacing ranging from 10 m at the surface to 457 m near the ocean bottom \citep{forget2015ecco}. Additionally, we use high-resolution precipitation estimates from the final run of the Integrated Multi-satellite Retrievals for GPM (IMERG), starting from the year 1998 \citep{huffman2015integrated}. The phase of the ENSO is determined using the Oceanic Ni\~{n}o Index (ONI) provided by the National Oceanic and Atmospheric Administration's (NOAA) Climate Prediction Center (CPC). The Dipole Mode Index (DMI) from NOAA is used to determine the phase of the Indian Ocean Dipole (IOD). The BSISO phases are from the bimodal ISO index developed by \citet{kikuchi2021boreal}.

\subsection*{Identifying MHWs}

To identify MHWs, we follow the definition proposed by \citet{hobday2016hierarchical}, which defines a MHW event as occurring when the SST exceeds the seasonally varying 90$\mathrm{^{th}}$ percentile  threshold for at least 5 consecutive days. MHW events separated by gaps of two days or less are treated as a single event. Climatological means and percentiles are calculated for each calendar day using an 11-day window centered on that day, with further smoothing applied through a 31-day moving average. An entire 42-year period of data (1982--2023) is used to compute the climatology. A similar methodology is also used to compute daily anomalies for other fields presented in this study. To investigate the drivers and impacts of MHWs, we remove the influence of anthropogenic warming from individual MHWs \citep{jacox2019marine,amaya2023marine,smith2025baseline} by detrending the SST data at each grid point, removing the linear trend over the 42-year period. More discussion on the relative merits of fixed and varying SST thresholds (to account for changes in the mean SST state) for identifying MHWs can be found in \citet{smith2025baseline}.

\subsection*{MHW budget analysis using ECCO reanalysis}

To identify the atmospheric and oceanic processes driving MHWs, we examine the mixed layer heat budget during these extreme warming events. We use NASA's Estimating the Circulation and Climate of the Ocean (ECCO V4r4) reanalysis data, which provides complete budget closure---unlike other reanalysis datasets---and whose representation of MHWs is generally consistent with observations \citep{sala2025leading}. Recent studies have highlighted the extension of MHWs beyond the surface into the ocean interior \citep{elzahaby2019observational, zhang2023vertical}, and thus, the mixed layer depth-averaged budget offers a more comprehensive perspective compared to surface-level analyses.

The temperature tendency of the mixed layer is governed by \citep{moisan1998seasonal}, 
\begin{align}
\frac{\partial T_m}{\partial t} = 
\underbrace{\frac{Q_{net}}{\rho C_p H} - \frac{Q_{SW}(z=-H)}{\rho C_p H}}_\text{net surface heat flux} 
\underbrace{- \frac{1}{H}  \int_{-H}^{0} \nabla(\!{\bf u} \!\cdot  T_{m}) \,dz}_\text{advection}
\underbrace{- \frac{1}{H}  \int_{-H}^{0} \nabla \!\cdot {F}_{\text{diff}} \,dz}_\text{diffusion}
\label{eqn:budget}
\end{align}
where, $T_m$ and H are the mixed layer temperature and mixed layer depth, respectively, and $\rho = 1029~\mathrm{kg~m^{-3}}$ and  $C_p = 3994~\mathrm{J~kg^{-1}~K^{-1}}$ are the density and heat capacity of seawater. The first term on the right-hand side represents the net surface heat flux, which includes net shortwave, net longwave, latent heat, and sensible heat fluxes, excluding the shortwave radiation that penetrates through the bottom of the mixed layer. The last two terms represent contributions from advection (both horizontal and vertical) and diffusion (both horizontal and vertical). These terms are vertically integrated from the surface (z = 0) to the mixed layer depth (z = -H).

\subsection*{WRF model and experiments}

To validate the relationship between MHWs and precipitation identified from satellite and reanalysis data, we conducted case studies using the Weather Research and Forecasting (WRF) model to investigate the impact of warm SST anomalies associated with MHWs on atmospheric circulation. WRF is a mesoscale numerical weather prediction system widely used to study high-impact weather systems across various regions globally \citep{powers2017weather}. The model was run with a horizontal grid spacing of $0.25^\circ \times 0.25^\circ$ and 40 vertical levels, with a time step of 60 seconds. Initial and boundary conditions are from the ERA5 dataset, while SST is from the OISST dataset. The Morrison 2-moment scheme \citep{morrison2009impact} is used to resolve the microphysics; the updated version of the Kain-Fritsch convection scheme \citep{kain2004kain} is used for cumulus parameterization; the Mellor-Yamada-Nakanishi-Niino 2.5-order closure scheme \citep{nakanishi2004improved, nakanishi2009development} is used for the planetary boundary layer and the surface layer; the Rapid Radiation Transfer Model for GCMs (RRTMG; \cite{iacono2008radiative}) is used for longwave and shortwave radiation transfer through the atmosphere; the Noah land surface model is used for the land surface processes \citep{tewari2004implementation}. 

Two MHW events are chosen for case studies, one in the eastern Arabian Sea and the other in the northern Arabian Sea. For each MHW event, the WRF model is run for two scenarios: the \textit{Main Run}, which is forced with daily-varying SST that includes the MHW-associated SST anomalies (SSTA), and the \textit{Clim Run}, which used daily climatological SST that excludes the SSTA associated with the MHW. The difference between these two simulations (\textit{Main Run - Clim Run}) isolates the impact of the MHW. Note that in the \textit{Clim Run}, only the initial and boundary conditions for SST were set to its climatological values, while the remaining initial and boundary conditions were kept the same as those in the \textit{Main Run}. Although retaining actual atmospheric conditions in the \textit{Clim Run} may introduce some biases, this approach avoids potential spin-up issues that can arise from initializing all variables with their climatological values. This setup allows the difference between the two runs to effectively capture the impact of MHW-related SST anomalies on precipitation.

\subsection*{Assessing the risk of extreme precipitation}

To assess how the risk of extreme precipitation during MHWs compares to that during non-MHW days, we compute the relative change in the risk of extreme precipitation between these two periods. The risk associated with MHWs is defined as $\text{Risk}_{\text{MHW}} = \frac{\mathcal{P}(P_{95} \mid P_{\text{MHW}})}{P_{\text{MHW}}}$, representing the conditional probability of occurrence of extreme precipitation during MHWs. Here, $\mathcal{P}(P_{95} \mid P_{\text{MHW}})$ is the proportion of days with extreme precipitation---defined as exceeding the 95$\mathrm{^{th}}$ percentile of JJAS rainfall---occurring within the first 5 days following the peak of a MHW, the period when most precipitation typically occurs. All other JJAS days are considered non-MHW days and are used to compute the corresponding baseline risk, $\text{Risk}_\text{non-MHW}$. The relative change in risk during MHWs is then $\Delta \text{Risk}_{\text{MHW}} = \frac{\text{Risk}_{\text{MHW}} - \text{Risk}_{\text{non-MHW}}}{\text{Risk}_{\text{non-MHW}}}$. 

Similarly, to assess the risk of extreme precipitation during BSISOs with and without MHWs, we define relative change in risk as $\frac{\text{Risk}_{\text{BSISO with MHW}} - \text{Risk}_{\text{BSISO without MHW}}}{\text{Risk}_{\text{BSISO without MHW}}}$. Here,  $\text{Risk}_{\text{BSISO with MHW}} = \frac{\mathcal{P}(P_{95} \mid P_{\text{BSISO with MHW}})}{P_{\text{BSISO with MHW}}}$, represents the proportion of days with extreme precipitation occurring during BSISOs with MHWs---specifically, within 5 days of the SST peak of the BSISO event---relative to the total number of such days at each grid point.

\section{Arabian Sea: Emerging hotspots of marine heatwaves}

\begin{figure}
    \centering
    \includegraphics[width=\textwidth]{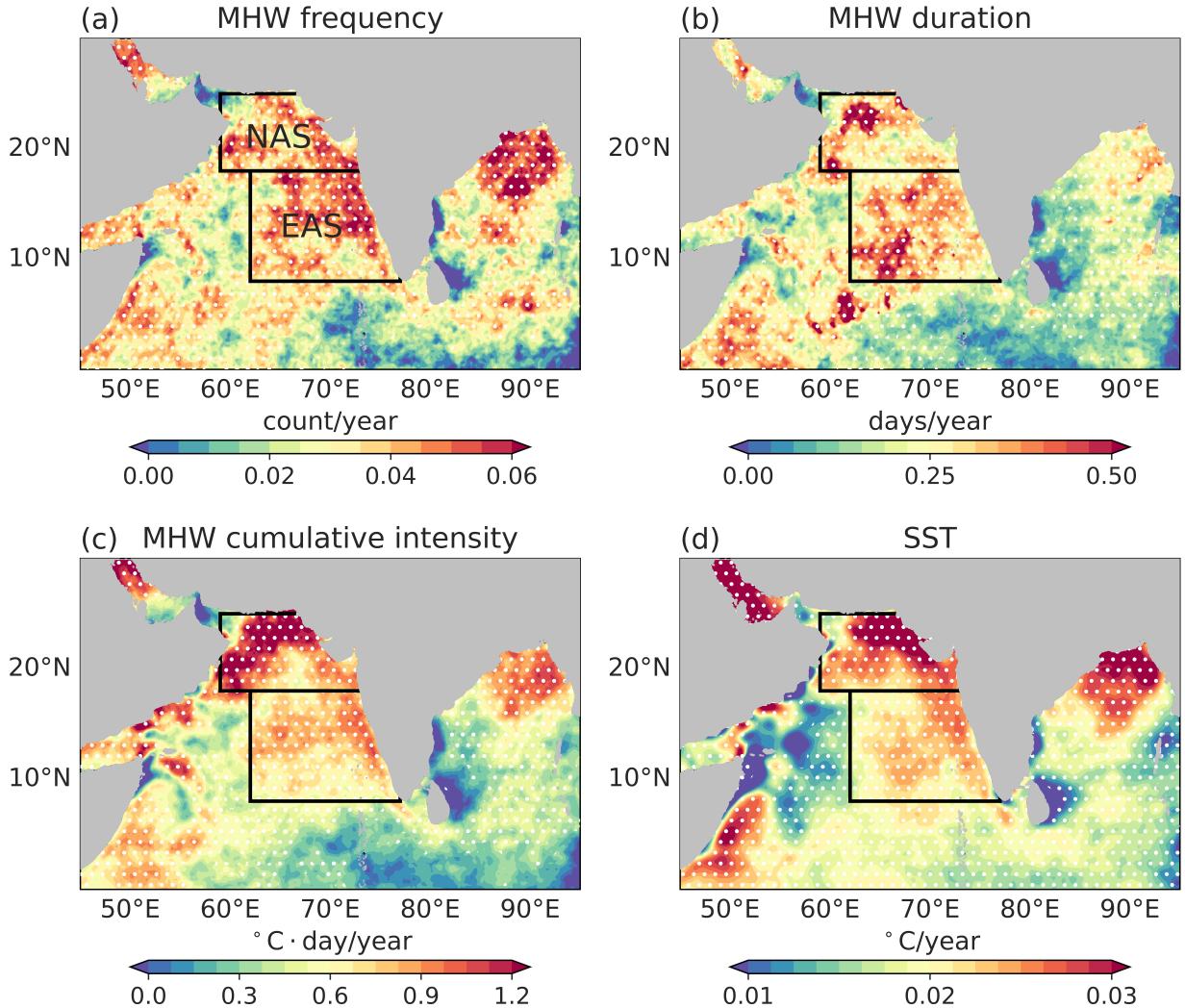}
    \caption{Linear trends of (a) MHW frequency, (b) MHW duration, (c) MHW cumulative intensity, and (d) SST over the North Indian Ocean during boreal summer monsoon season (June--September) from 1982 to 2023, based on satellite-observed daily SST data. Stippled regions indicate trends that are significant at the 95\% confidence level. Black boxes denote the regions with a large increase in MHWs over the Arabian Sea: the eastern Arabian Sea (EAS; 8$^\circ$N-18$^\circ$N, 62$^\circ$E-77$^\circ$E) and northern Arabian Sea (NAS; 18$^\circ$N-25$^\circ$N, 59$^\circ$E-73$^\circ$E).}
    \label{fig:MhwTrends}
\end{figure}

The eastern and northern Arabian Sea, along with the northern Bay of Bengal and western equatorial region off the Somali coast, show larger increases in the frequency, duration, and cumulative intensity of MHWs during boreal summer, compared to other areas of the North Indian Ocean (Figs. \ref{fig:MhwTrends}a-c; see also Supplementary Fig. S1). In this study, we focus on the eastern and northern Arabian Sea regions, which have emerged as prominent MHW hotspots in the Indian Ocean since 1982. We divide the Arabian Sea into northern and eastern subregions, to examine whether similar drivers are at play in both regions and to assess their respective impacts. 

The spatial pattern of increased MHW activity closely mirrors the SST warming trend from 1982--2023, with the most rapid warming occurring in the MHW hotspot regions (compare Figs. \ref{fig:MhwTrends}a-c with \ref{fig:MhwTrends}d). The sensitivity of these trends to the thresholds used to define MHWs is tested in Supplementary Fig. S2, which shows little difference. While anthropogenic warming---estimated from the multi-model ensemble mean of high resolution CMIP6 models---can explain a large fraction of the observed warming trend in the Arabian Sea, it cannot reproduce the observed spatial patterns and magnitude of the warming trend (Supplementary Fig. S3). For instance, the strong SST warming observed in the northern Arabian Sea is not captured in the multi-model mean trend of the CMIP6 simulations (Fig. S3a). The discrepancy between model simulations and observations suggests that internal climate variability plays a major role in shaping the observed warming patterns, or that the CMIP6 models, despite their high resolution, exhibit systematic biases.

\begin{figure}
    \centering
    \includegraphics[width=\textwidth]{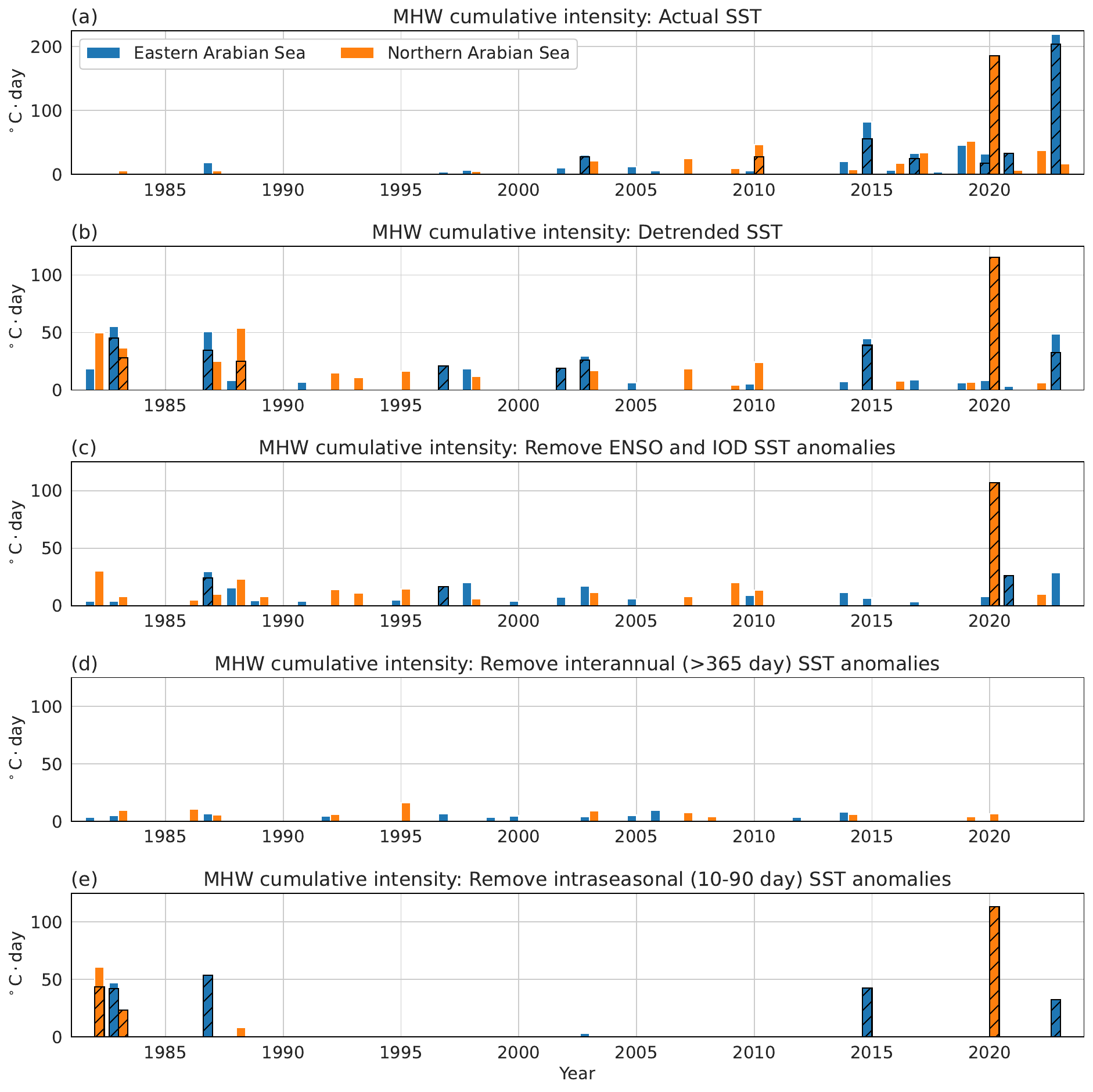}
    \caption{(a) Cumulative intensity of MHWs in the eastern and northern Arabian Sea, averaged for each summer (June--September), from 1982 to 2023. (b) Same as (a), but using detrended SST to identify MHWs. (c) as in (b), but with the influences of ENSO and IOD removed from the detrended SST (see text). (d,e) As in (b), but with (d) interannual (\textgreater 365 day) and (e) intraseasonal (10–90 day) SSTA removed from the detrended SST. Panels (c--e) use the same $90^{th}$ percentile threshold as in panel (b) to identify MHWs. Contributions from longer-lasting MHWs (duration \textgreater 20 days) are hatched.}
    \label{fig:MhwCumIntYearly}
\end{figure}

The increasing occurrence of MHWs in the eastern and northern Arabian Sea is also evident in the time series of their cumulative intensity, which provides an integrated measure of MHW activity (Fig. \ref{fig:MhwCumIntYearly}a). This upward trend is consistent with the rising SST in each region, which increases the likelihood of exceeding the fixed 90$^{\mathrm{th}}$ percentile threshold (1982--2023) used to identify MHWs. Although a fixed SST baseline is useful for evaluating the MHW ecological impacts \citep{smith2023biological, li2023marine}, it can also lead to a perpetual state of MHWs with continued ocean warming, which can distort assessments of both the drivers and impacts of MHWs \citep{jacox2019marine, amaya2023marine, farchadi2025marine, smith2025baseline}. To isolate the transient extreme warming events (i.e., MHWs) from the background warming trend, we can either apply a time-evolving baseline or detrend the SST data before identifying MHWs \citep{hobday2016hierarchical, amaya2023marine, jacox2019marine, jacox2020thermal, burger2022compound, li2023marine, smith2025baseline}. In this paper, we remove the linear trend from the SST data prior to identifying MHWs hereafter, thereby excluding the influence of anthropogenic warming as well as the 42-year trend that resulted from multi-decadal climate variability. The linear trend reasonably captures the overall increase in SST during the analysis period (Supplementary Fig. S3b). Compared to Fig. \ref{fig:MhwCumIntYearly}a, MHWs identified using detrended SST display clear interannual variability without a discernible long-term trend (Fig. \ref{fig:MhwCumIntYearly}b). 

\section{Drivers of marine heatwaves}

\begin{figure}
    \centering
    \includegraphics[width=\textwidth]{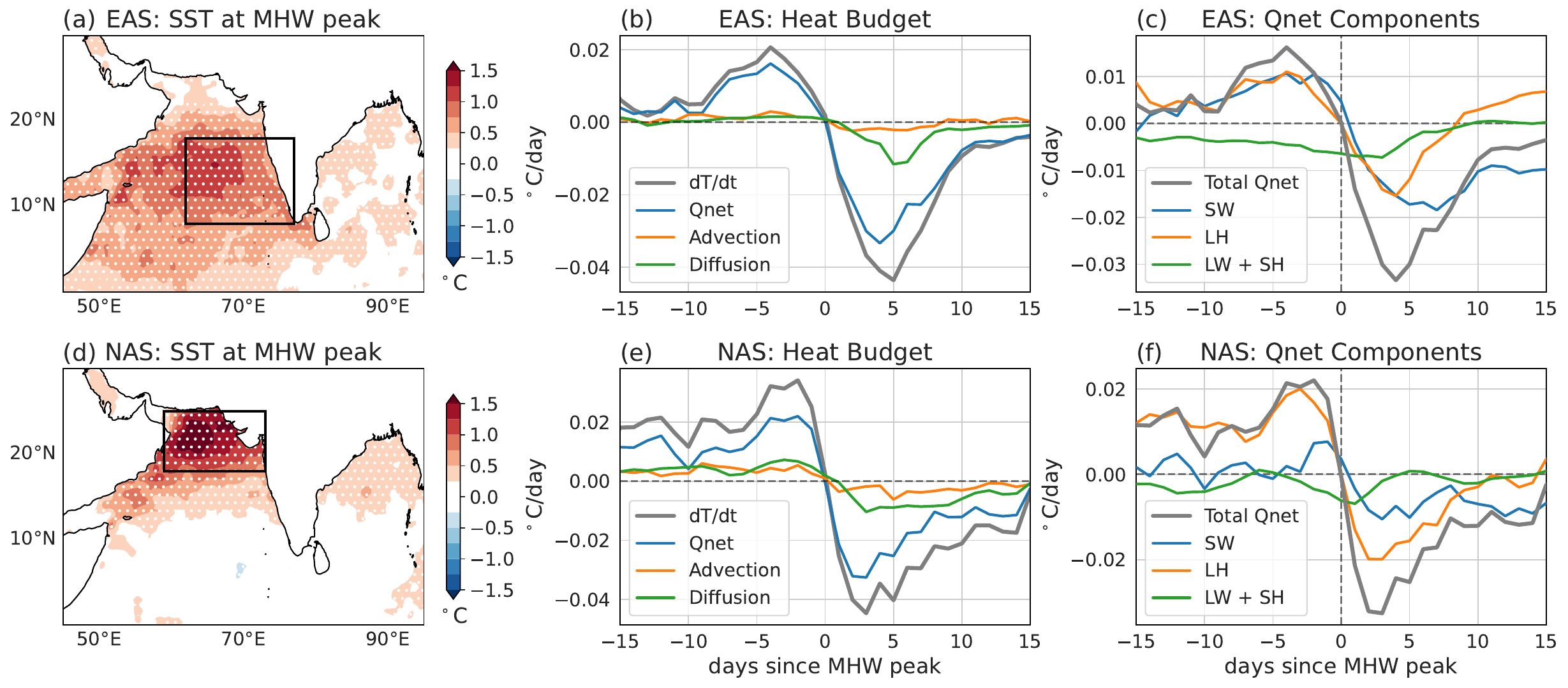}
    \caption{Composite of (a) SST anomalies on the day of MHW peak intensity, (b) mixed layer heat budget during MHWs, and (c) anomalous contributions from various components of net air-sea heat fluxes, for MHWs in the eastern Arabian Sea. Panels (d--f) are the same as (a--c), but for the northern Arabian Sea. SSTA is from OISST. Stippling indicates values significant at the 95 \% confidence level based on a t-test. Mixed layer heat budget is from ECCO reanalysis, where $Q_{net}$ represents the net air–sea flux, corrected for shortwave radiation exiting the mixed layer. Advection and diffusion terms include both horizontal and vertical contributions. The composite plots are centered on the peak intensity of each MHW (day 0), with negative days representing the developing phase and positive days representing the decaying phase of the MHWs.}
    \label{fig:SstBudget}
\end{figure}

MHWs in the Arabian Sea are associated with basin-wide SSTA (Figs. \ref{fig:SstBudget}a and d), suggesting the influence of large-scale processes in generating these events. During the Indian summer monsoon season, warm SSTAs in the Arabian Sea can arise due to  El Ni\~{n}o---both during its developing phase in summer and following its peak phase from the previous winter---as well as by the positive phase of the IOD \citep{xie2009indian, yang2007impact, klein1999remote, saji1999dipole}. Both factors likely increase the MHW frequency and intensity. Indeed, more MHWs are observed during El Ni\~{n}o and positive IOD compared to La Ni\~{n}a and negative IOD (Supplementary Fig. S4a). While ENSO and IOD might indirectly influence MHWs by modulating BSISO activities---an important driver of MHWs, as will be discussed below---both the frequency and intensity of BSISOs remain similar during El Ni\~{n}o and La Ni\~{n}a years, as well as across positive and negative IOD phases (Fig. S4b). This suggests that ENSO and IOD modulate MHWs primarily through their direct impacts on SSTA (Fig. S4c). 

To quantify the impacts of ENSO and IOD on June--September SSTA---and consequently on the year-to-year variability of boreal summer MHW activities---we first obtain the ENSO- and IOD-associated SSTAs using linear regression, a proven approach widely used in previous studies \citep{klein1999remote, oliver2018longer, marathe2021tropical, ashok2001impact, zhang2020effects}. These ENSO- and IOD-related SSTAs are then removed from the observed SSTAs before identifying MHW events. Since Arabian Sea SSTs are influenced by ENSO both during its developing phase in summer and after its peak in the preceding winter, we remove the effects of ENSO by accounting for both the concurrent (lag-0) and prior DJF Oceanic Ni\~{n}o Index (ONI). Similarly, the influence of the IOD is removed using the lag-0 Dipole Mode Index (DMI). After removing the effects of ENSO and the IOD, the cumulative intensity of MHWs---especially those lasting more than 20 days---decreases (Fig. \ref{fig:MhwCumIntYearly}c, hatched bars). However, nearly all MHWs still occur, and most short-lived events (lasting $\le$ 20 days) remain largely unaffected. We chose a 20-day threshold to separate the relatively few long-lasting MHWs from the majority of shorter events, which are more strongly influenced by intraseasonal variability (discussed below). The median duration of all MHWs in both the eastern and northern Arabian Sea is approximately 9 days, reflecting the predominance of short-lived events. Of the 32 MHWs in the eastern Arabian Sea, 25 are short-lived, while in the northern Arabian Sea, 25 out of 29 MHWs are short. The 20-day threshold therefore highlights the contrast between short and long events, and varying it by a few days does not change the results. When all SSTAs with periods longer than a year (which  includes ENSO and IOD effects) are removed, a substantial reduction in MHW activities is observed, and all long MHWs---including the unusually strong 2020 event in the northern Arabian Sea---disappear (Fig. \ref{fig:MhwCumIntYearly}d). This indicates that interannual SSTAs induced by factors other than ENSO and IOD also have significant influence on MHW activities. Notably, when intraseasonal (10--90 day) SSTAs is removed, all short MHWs vanish, whereas most long MHWs (such as the 2020 northern Arabian Sea event) remain intact---except for some weaker ones which are either significantly reduced or disappear (compare Figs. \ref{fig:MhwCumIntYearly}b and \ref{fig:MhwCumIntYearly}e). 

The above results indicate that long MHWs in the eastern and northern Arabian Sea during the Indian summer monsoon season arise from interannual SSTAs driven by large-scale climate variability (e.g., El Ni\~{n}o and positive IOD), with intraseasonal SSTAs either enhancing or weakening some of these events. This explains why they almost completely vanish when interannual SSTAs are removed but are weakly affected by the removal of intraseasonal SSTAs. Interestingly, the two strongest long MHWs in terms of cumulative intensity, occurring in 2020 and 2023 (Fig. \ref{fig:MhwCumIntYearly}a), appear to be associated with different factors: the 2023 MHW largely disappears when the long-term SST trend is removed (Fig. \ref{fig:MhwCumIntYearly}b), suggesting a strong contribution from ongoing warming, which may also have contributed to unprecedented MHWs in other regions in 2023 \citep{dong2025record}, whereas the 2020 MHW is largely unaffected by the long-term trend and only reduces when interannual variability is removed (Figs. \ref{fig:MhwCumIntYearly}b-d), indicating that it is primarily caused by interannual SSTAs independent of long-term warming. While it is well known that ENSO and IOD affect Indian Ocean SST, climate variability that induces strong interannual SSTA in the Arabian Sea but is unrelated to ENSO and IOD remains less studied. We have also examined the effects of other known climate modes (e.g., the North Atlantic Oscillation, Atlantic Multidecadal Variability, and Pacific Decadal Oscillation), and none of them can adequately account for the observed interannual variations (Supplementary Fig. S5), highlighting the need for further investigation. In contrast, short MHWs are largely controlled by intraseasonal SST variability, as evidenced by their disappearance when intraseasonal SSTAs are removed, with warm interannual SSTAs acting to enhance them. Given that intraseasonal SST variability is essential for causing individual short MHW events---which constitute the majority of observed events---we now focus on understanding the drivers of intraseasonal SSTAs associated with MHWs.

One of the potential drivers of these intraseasonal SSTAs is the BSISOs. The Arabian Sea is strongly influenced by BSISOs during the Indian summer monsoon season (June--September) \citep{kikuchi2021boreal}. The wind and convection anomalies associated with BSISOs can induce intraseasonal SSTAs \citep{sengupta2001coherent}, with some phases of the BSISOs favoring SST warming (Supplementary Fig. S6). In both the eastern and northern Arabian Sea, the onset and peak of MHWs typically follow the suppressed convection phase of BSISOs (not shown), which is characterized by enhanced surface heat fluxes and subsequent SST warming (Fig. S6)---conditions that favor MHW development.

\begin{figure}
    \centering
    \includegraphics[width=\textwidth]{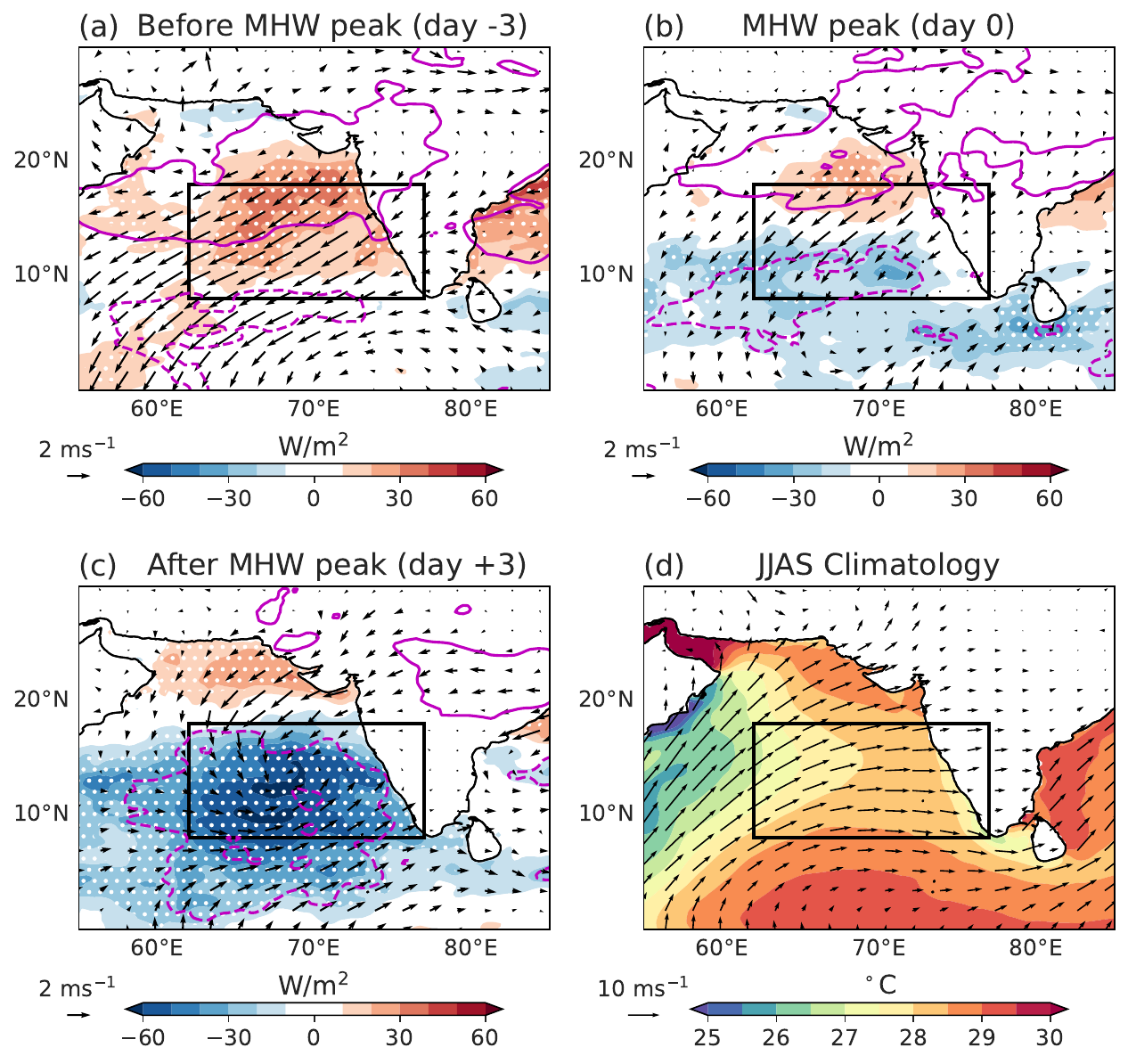}
    \caption{Evolution of anomalous net air-sea heat flux (color), outgoing longwave radiation (OLR; purple contours---positive solid and negative dashed; units: $\mathrm{Wm^{-2}}$), a proxy for tropical atmosphere deep convection, and surface winds (vectors) associated with MHWs in the eastern Arabian Sea (black box). Panels show (a) day -3, (b) day 0, and (c) day 3, with day 0 marking the MHW peak. Stippling indicates values significant at the 95 \% confidence level based on a t-test. Panel (d) shows the climatological boreal summer (JJAS) mean SST and surface winds over the Arabian Sea. Fluxes are from ERA5 reanalysis, SST is from OISST, and surface winds are from the Cross-Calibrated Multi-Platform (CCMP) Ocean Surface Wind Vector Analyses.}
    \label{fig:QnetEvoln}
\end{figure}

Indeed, mixed layer heat budget analysis using the Estimating the Circulation and Climate of the Ocean (ECCO) data confirms the importance of net surface heat flux in driving the SSTA during MHWs (Figs. \ref{fig:SstBudget}b and e). In the eastern Arabian Sea, during the developing phase of MHWs (before day 0, where day 0 refers to the peak of the MHW), net positive surface heat flux is the dominant driver for increasing the SST (Fig. \ref{fig:SstBudget}b), mainly through shortwave radiation and latent heat flux, which contribute comparably (Fig. \ref{fig:SstBudget}c). Advection and diffusion also have some contributions (Fig. \ref{fig:SstBudget}b), with horizontal advection and vertical mixing dominating their respective terms (not shown). During the decaying period (after the MHW peak), both the latent and shortwave fluxes become negative; together with advection and diffusion, they lead to cooling of SST. These results align with previous SST budget analyses in the region (though not focused specifically on MHWs), which indicate that intraseasonal SST variability is primarily driven by air–sea heat fluxes, with additional contributions from advection and mixing \citep{li2016intraseasonal}. The strong monsoonal winds over the Arabian Sea is an important factor in affecting vertical mixing \citep{shenoi2002differences}, and the weakening and strengthening of these winds during the developing and decaying phases of MHWs (as will be discussed in Fig. \ref{fig:QnetEvoln}) can contribute to the anomalous warming and cooling of SST (Fig. \ref{fig:SstBudget}b). The situation in the northern Arabian Sea is similar, except that advection and weakened mixing contribute more to the SST warming during the MHW developing phase (Figs \ref{fig:SstBudget}e--f). These changes in air-sea heat fluxes typically occur over a ~20-day period, emphasizing the role of intraseasonal variability in driving MHWs.

The spatio-temporal evolutions of net surface heat flux, convection---represented by outgoing longwave radiation (OLR)---and surface wind composites for MHW events provide further insights into the relevant processes. Before the MHW reaches its peak intensity, the eastern Arabian Sea experiences conditions similar to those during the suppressed convection phases of BSISOs (phases 1 and 2 in Supplementary Fig. S6). These phases are characterized by clear skies, as indicated by positive OLR anomalies (Figs. \ref{fig:QnetEvoln}a and S6), leading to enhanced shortwave radiation at the ocean surface. Concurrently, northeasterly surface wind anomalies weaken the prevailing southwesterly monsoon winds (Fig. \ref{fig:QnetEvoln}d), reducing total wind speed and thereby suppressing evaporation and latent heat loss. Together, these BSISO-modulated anomalies in wind and cloud cover create conditions conducive to SST warming and the subsequent development of MHWs. A similar situation prevails in the northern Arabian Sea as well (Supplementary Fig. S7a), where the conditions resemble those of BSISO phases that promote northern Arabian Sea warming (phases 4 and 5 in Supplementary Fig. S6), characterized by enhanced net air–sea heat fluxes. As similar drivers are present in both regions, the discussion focuses primarily on the eastern Arabian Sea, while key results from the northern Arabian Sea are included in the main text and additional results are provided in the Supplementary Material. Any notable differences between the two regions are highlighted where relevant.

Following the MHW peak, a low-level cyclonic wind anomaly and increased convection (indicated by negative OLR anomalies) develop over the eastern Arabian Sea, which in turn reduce downward shortwave radiation into the ocean (Fig. \ref{fig:QnetEvoln}c). Surface winds, particularly in the southern part of the eastern Arabian Sea, become southwesterly, strengthening the prevailing southwest monsoon flow (Fig. \ref{fig:QnetEvoln}d) and thereby enhancing evaporation and latent heat loss to the atmosphere. These changes contribute to the subsequent cooling of the SST. While most studies on BSISOs focus on the 25--90 day periods \citep{kikuchi2021boreal}, our analysis shows that the quasi-biweekly (10--25 day) modes \citep{goswami2001intraseasonal,karmakar2017space} also contribute substantially to MHWs, particularly at the MHW peak (Supplementary Fig. S8). A similar low-level cyclonic circulation also develops in the northern Arabian Sea following the MHW peak (Supplementary Fig. S7b), indicating that MHWs exert feedbacks onto the atmosphere in both regions, leading to the formation of low-level cyclones. Notably, some northern Arabian Sea MHWs, although not all, are associated with northward-propagating BSISOs that initially induce MHWs in the eastern Arabian Sea. Of the 29 MHWs observed during the JJAS season in the northern Arabian Sea, 5 were preceded by eastern Arabian Sea MHWs within 15 days, and 7 within 30 days.

\section{Impact of marine heatwaves on weather and monsoon}

\subsection{Effects on regional atmospheric circulation and weather} 

\begin{figure}
    \centering
    \includegraphics[width=\textwidth]{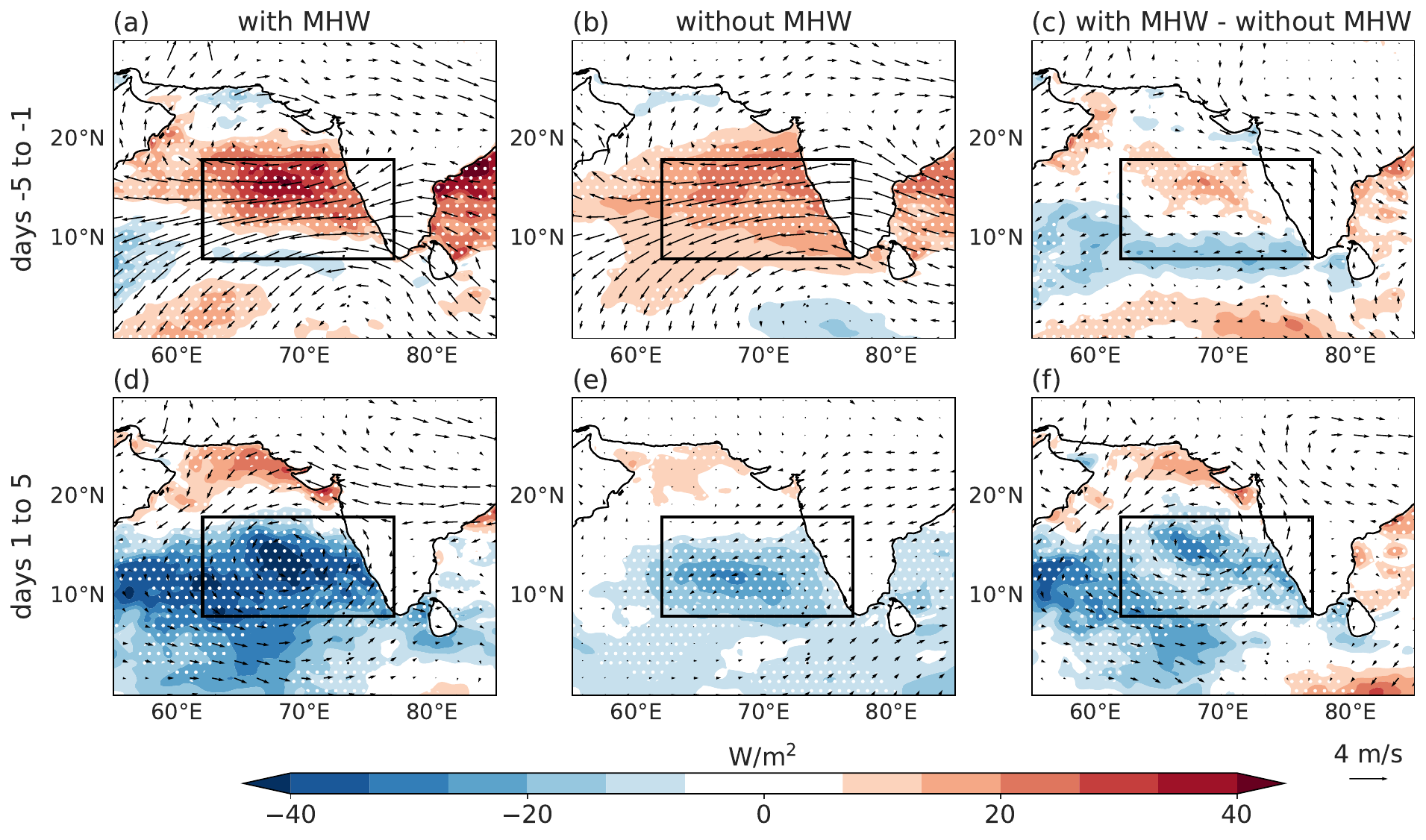}
    \caption{The anomalous net air-sea heat flux (color) and 850 hPa winds (vectors) for BSISOs occurring (a,d) with MHWs and (b,e) without MHWs, along with their differences (c,f). BSISOs are centered around their SST maxima, with day 0 marking the peak SST. The SST maximum is considered over the eastern Arabian Sea (black box). Panels (a-c) display conditions prior to the SST maxima (days -5 to -1), while panels (d-f) show conditions after the SST peak (days 1-5). Stippling indicates values significant at the 95 \% confidence level based on a t-test. Air-sea fluxes and winds are from ERA5, while SST is from OISST dataset.}
    \label{fig:BsisoMhwDiff}
\end{figure}

While BSISOs drive MHWs, a key question arises: Are MHWs merely an oceanic response to intense BSISO events, or do they also actively impact atmospheric circulation and weather patterns? As discussed earlier, a low-level cyclonic circulation anomaly develops following the MHW peak (Fig. \ref{fig:QnetEvoln}c), suggesting that the warm SSTAs associated with MHWs are likely inducing the ciruclation anomaly. To confirm this hypothesis and answer the above question, we analyze the large-scale conditions associated with BSISO events that occur with and without MHWs, focusing on the eastern Arabian Sea (Fig. \ref{fig:BsisoMhwDiff}). We define a BSISO event as starting at phase 6 and ending at phase 5 in the eastern Arabian Sea, and starting at phase 1 and ending at phase 8 in the northern Arabian Sea. These phase boundaries correspond to SST minima in each region, thereby encompassing the peak phase of MHWs within the same BSISO cycle. A MHW is considered to occur during a BSISO event if it takes place at any time within the corresponding BSISO cycle; otherwise, the event is classified as a BSISO without MHW. To allow a direct comparison with MHWs, we composite the BSISO around the SST peak (t = 0) rather than the peak amplitude of the BSISO index or OLR. By aligning the BSISO to the SST peak, the MHW peak (when present) coincides with the SST peak during the BSISO cycle, so that ``BSISO with MHWs" refers specifically to conditions near the SST peak and does not necessarily represent the entire BSISO cycle.

Compared to BSISO events without MHWs, those accompanied by MHWs exhibit notably stronger net surface heat fluxes before and after the SST peak (Fig. \ref{fig:BsisoMhwDiff}). Low-level atmospheric winds show little difference between the two scenarios prior to the SST peak (Figs. \ref{fig:BsisoMhwDiff}a--c). Strikingly, a pronounced cyclonic circulation develops following the SST peak for BSISOs with MHWs, which is not observed in the BSISOs without MHWs (Figs. \ref{fig:BsisoMhwDiff}d-f). The cyclonic wind anomalies are similar to that of MHW composites (Fig. \ref{fig:QnetEvoln}c), supporting the hypothesis that the low-level cyclonic circulation is driven by strong SST anomalies associated with MHWs. 

To confirm the role of MHWs in generating the cyclonic circulation in low-level atmosphere, we carry out two parallel numerical experiments using the WRF model (see subsection \textit{WRF model and experiments}). These experiments are designed to assess the impact of a strong MHW event that began on June 1, 2014 and reached its peak intensity on June 8, 2014 in the eastern Arabian Sea. We chose the 2014 short MHW event instead of other long MHW (despite the latter having a larger cumulative intensity) because it is more representative of the majority of MHWs in this region. This MHW event lasted 9 days, similar to the median duration of MHWs in the EAS, and had a cumulative intensity of $7.44^\circ\text{C·day}$, comparable to the climatological median for short MHWs ($5.44^\circ\text{C·day}$), and was associated with stronger but qualitatively similar background conditions compared to other MHWs in the region (Supplementary Fig. S9). The WRF model was initialized on June 5, 2014, with initial and boundary conditions supplied from the ERA5 dataset, while SST was from the OISST dataset. In the experiment \textit{Main Run}, the WRF model is forced with observed daily SST data from OISST; in the experiment \textit{Clim Run}, WRF model is forced with daily climatological SST, and their difference (\textit{Main Run - Clim Run}), isolates the impact of the MHW.

\begin{figure}
    \centering
    \includegraphics[width=\textwidth]{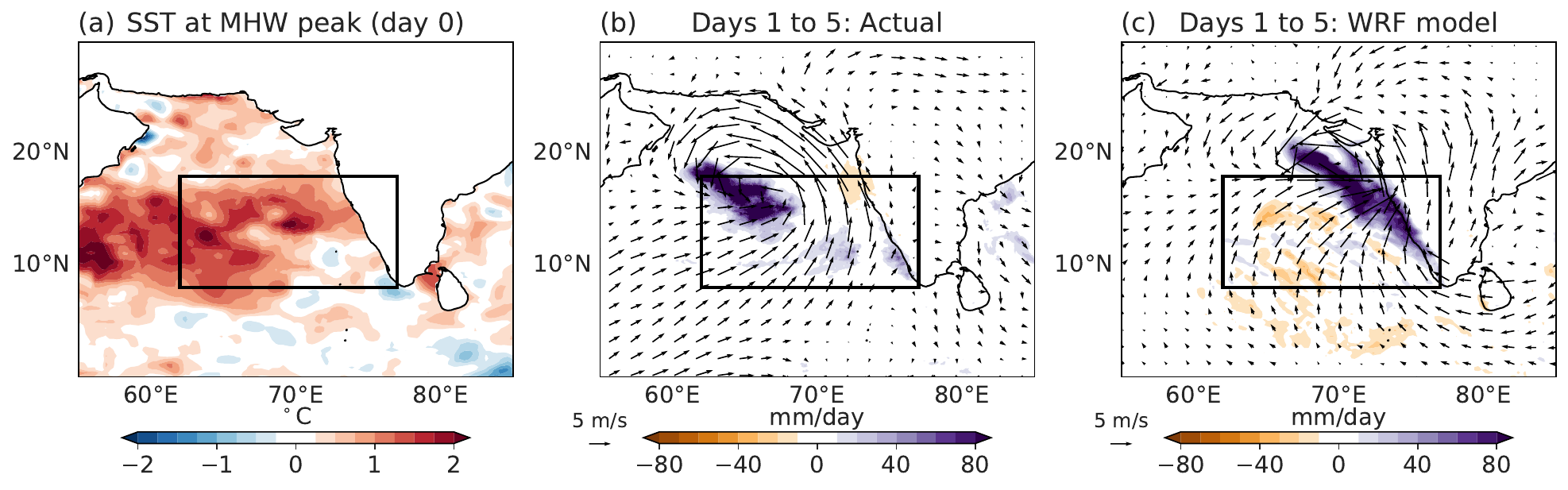}
    \caption{Case study of a MHW that began on June 1, 2014, using the WRF model. Panel (a) shows the detended SSTA (OISST) on the day when MHW reached its peak intensity (June 8, 2014). Panel (b) shows observed precipitation anomalies (color; IMERG) and surface wind anomalies (vectors; CCMP) during the decay phase (up to 5 days following the MHW peak).  Panel (c) is the same as (b) but shows output from the WRF Main Run -- Clim Run (see text).}
    \label{fig:WrfPrecip}
\end{figure}

In the observations, a cyclonic wind pattern and anomalous rainfall develop following the warm SSTA associated with the MHW (Fig. \ref{fig:WrfPrecip}a-b), closely resembling the patterns seen in the MHW composites (Figs. \ref{fig:QnetEvoln}c, \ref{fig:BsisoMhwDiff}f and \ref{fig:PrecipEAS}a). The difference between the two WRF scenarios (\textit{Main Run - Clim Run}) successfully simulates a low-level cyclonic circulation that develops shortly after the MHW peak, leading to precipitation over the Arabian Sea and southern Western Ghats (Fig. \ref{fig:WrfPrecip}c). We note that the WRF-simulated cyclone and its associated precipitation propagate faster than observed, resulting in a cyclonic anomaly that is both more northward and closer to the western coastline of India, with the precipitation anomaly extending farther north along the coast compared to the observations (Figs. \ref{fig:WrfPrecip}b–c). This discrepancy likely arises because the model experiments are forced by SST, which suppresses air–sea coupling processes that typically slow the propagation of atmospheric circulation and precipitation \citep{fu2003coupling}, but remote influences from other regions, WRF model biases, or internal variability may also play a role. Regardless, these WRF experiments show that the cyclonic circulation and associated precipitation develop only when the strong SST warming associated with MHWs is present, demonstrating that this is a distinct response caused by MHWs rather than an amplification of the BSISO-induced response.

\subsection{Impact on monsoon rainfall and extreme precipitation over land}

\begin{figure}
    \centering
    \includegraphics[width=.95\textwidth]{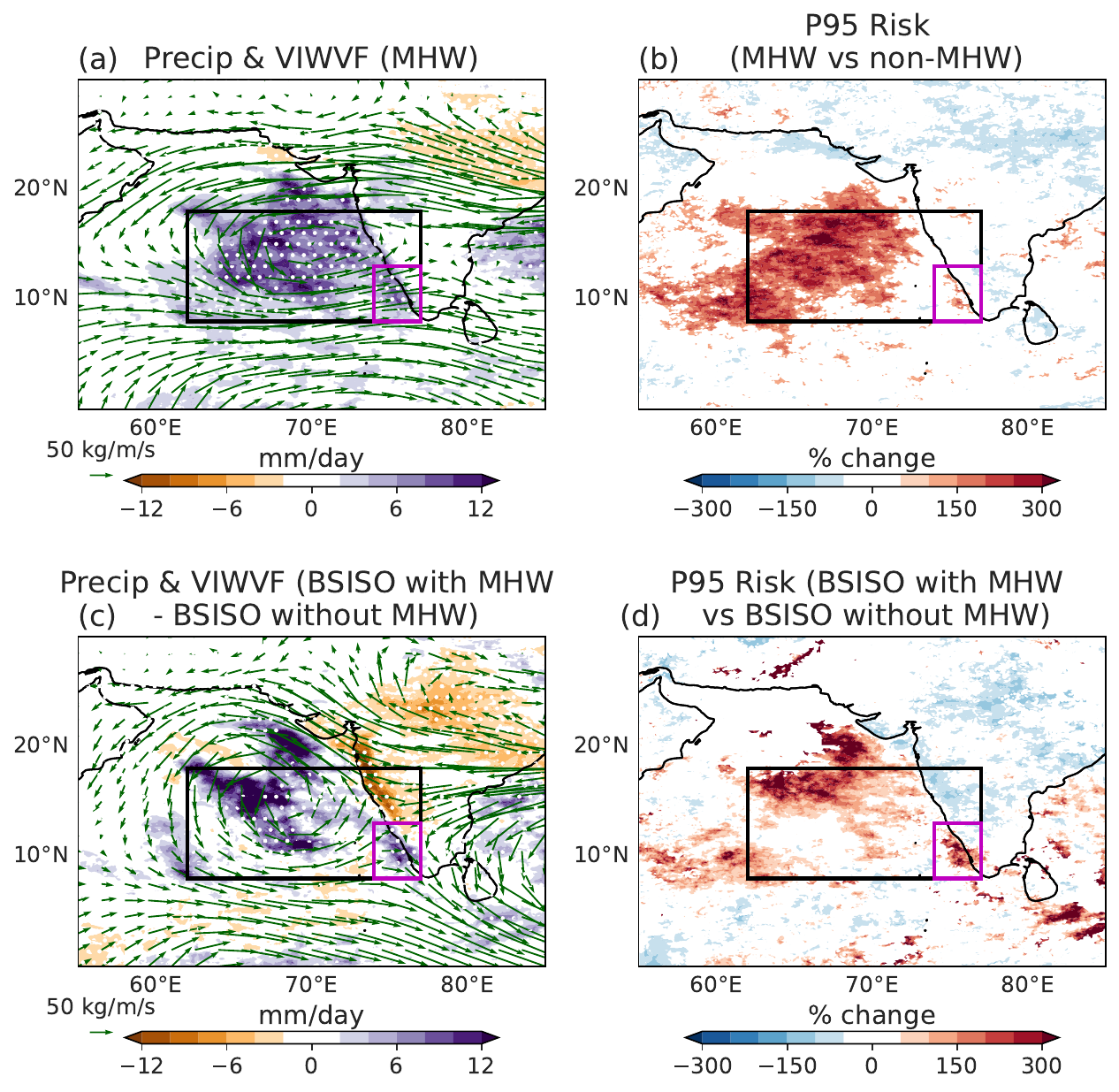}
    \caption{Anomalous precipitation and changes in extreme precipitation risk associated with MHWs in the eastern Arabian Sea. (a) Composite of anomalous precipitation (color) and vertically integrated water vapor flux (VIWVF; green vectors) during the decay phase of MHWs (days 1–5 after MHW peak). Stippling indicates values significant at the 95 \% confidence level based on a t-test. (b) Relative change in the risk of extreme precipitation (defined as exceeding the 95$\mathrm{^{th}}$ percentile of JJAS rainfall) during MHWs relative to non-MHW days, shown only where risk of precipitation exceeding the 95th percentile is $\geq 5\%$ for non-MHWs or $\geq 10\%$ for MHWs (double the baseline). (c) Same as (a), but showing the difference between BSISOs that occur with MHWs and those without MHWs. (d) Same as (b) but comparing the change in risk of extreme precipitation during BSISOs with MHWs compared to those without. Precipitation (IMERG) and water vapor fluxes (ERA5) are shown for 1998--2023, the common period between the two datasets. Anomalies are computed using daily climatologies. The purple box marks the area near Kerala in southern India with an increased risk of extreme precipitation.
    }
    \label{fig:PrecipEAS}
\end{figure}

In the eastern Arabian Sea, the MHW-induced cyclonic circulation and precipitation extend from the ocean onto land, causing excess rainfall along the southwest coast of India. Westerly or southwesterly wind anomalies on the equatorward side of the cyclone enhance moisture transport from the Arabian Sea to the southern part of the Western Ghats, as shown by composites of MHW events (Fig. \ref{fig:PrecipEAS}a). The increased moisture transport intensifies rainfall and raises the risk of extreme precipitation events (defined as exceeding the $95^{th}$ percentile of JJAS precipitation) along the southwest coast of India, particularly near Kerala (purple box in Fig. \ref{fig:PrecipEAS}). In contrast, winds on the poleward side of the cyclone are directed from land to ocean, resulting in reduced or unchanged precipitation over the northern Western Ghats, thereby creating a meridional contrast in rainfall along the west coast of India. This contrast is also evident in the WRF model simulations, where precipitation is enhanced on the equatorward side of the cyclonic anomaly (Fig. \ref{fig:WrfPrecip}c).

This south-north contrast in precipitation is more evident when comparing BSISO events with and without MHWs, revealing a clear wet-dry dipole pattern along the Western Ghats (Fig. \ref{fig:PrecipEAS}c). While BSISOs are known to increase the risk of extreme precipitation \citep{lee2017subseasonal}, MHWs further amplify this risk---enhancing the risk of extreme rainfall in southwestern India while reducing it over the northern Western Ghats and much of the Indian subcontinent (Fig. \ref{fig:PrecipEAS}d). For instance, Kerala faces a 2--3 times higher risk during BSISOs accompanied by MHWs compared to BSISOs without MHWs.

\begin{figure}
    \centering
    \includegraphics[width=.95\textwidth]{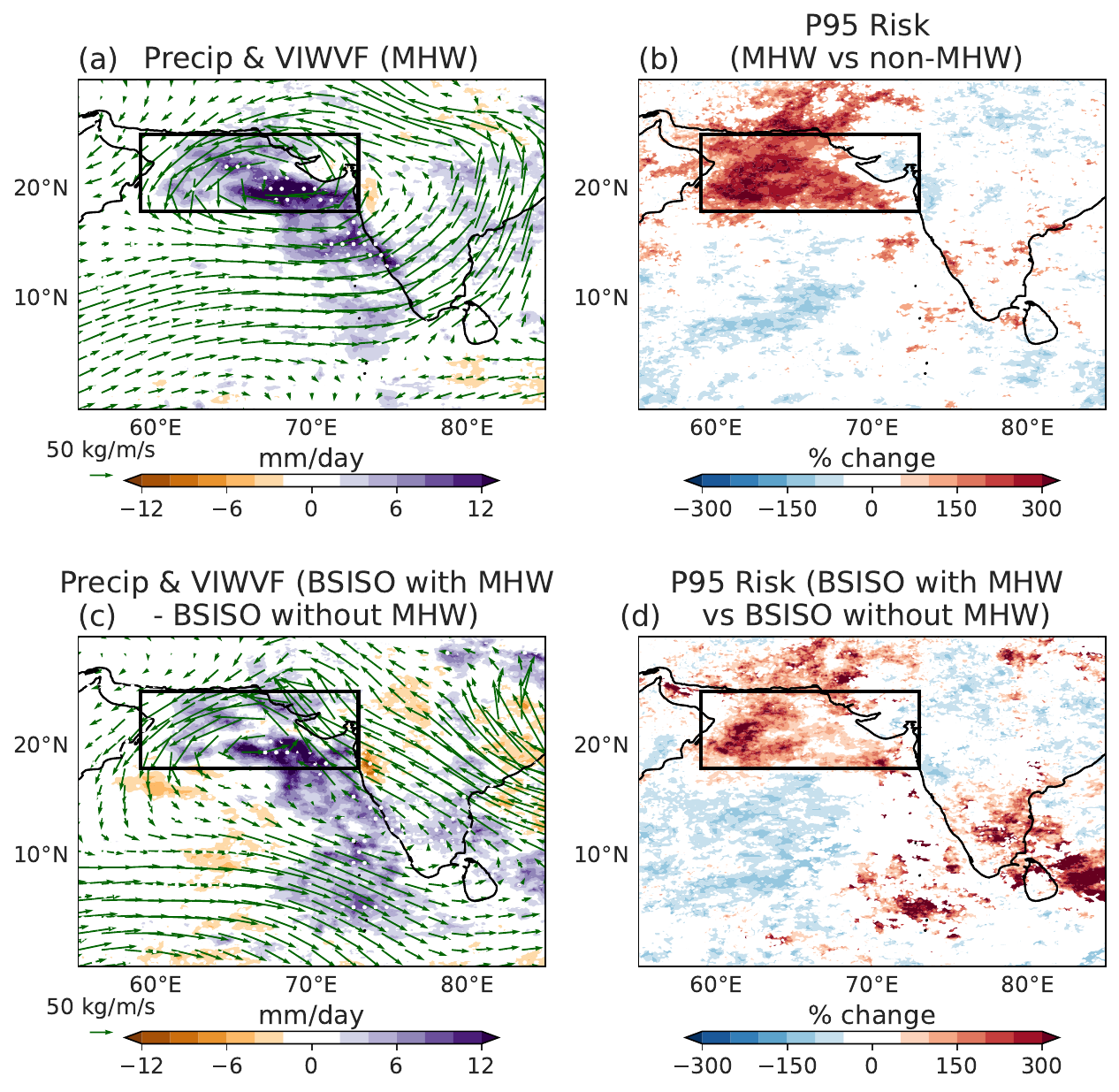}
    \caption{Same as Fig. \ref{fig:PrecipEAS} but for MHWs in the northern Arabian Sea.
    }
    \label{fig:PrecipNAS}
\end{figure}

The duration of a MHW event also influences precipitation patterns, with shorter events (lasting $\le$ 20 days) and longer events (\textgreater 20 days) having distinct impacts (Supplementary Fig. S10). Only shorter MHWs are associated with the development of low-level cyclonic winds and increased precipitation over the Arabian Sea, accompanied by contrasting precipitation along the western coast of India that occurs within a few days of the MHW peak (Fig. S10a). No such comparable changes occur during long MHWs (Fig. S10b). This likely reflects the stronger influence of intraseasonal anomalies (mainly BSISOs) on short MHWs, allowing a rapid development of circulation and precipitation response. A similar analysis for longer MHWs shows no consistent response; the lower frequency of long events, their more variable durations, and the potential influence of various interannual processes, which may produce differing responses across events, make the composite response of long MHWs difficult to interpret (Fig. S10b). Our analysis remains robust even when the 20-day threshold is varied by a few days. As most MHWs are short (e.g., 25 out of 32 in the eastern Arabian Sea), the composite MHW precipitation pattern primarily reflects the influence of these shorter events (compare Fig. \ref{fig:PrecipEAS}a and Supplementary Fig. S10).

In the northern Arabian Sea, MHW-induced low-level wind anomalies are also characterized by cyclonic circulation and accompanied by increased precipitation following the MHW peak (Figs. \ref{fig:PrecipNAS}a and c). During the decay phase of northern Arabian Sea MHWs, onshore winds associated with the low-level cyclone enhance moisture convergence and intensify rainfall along the central west coast of India (Fig. \ref{fig:PrecipNAS}a), further north than the region affected in Fig. \ref{fig:PrecipEAS}a. In addition, enhanced rainfall associated with the cyclonic anomaly is also observed in northwestern India and Pakistan, with MHWs increasing the risk of extreme precipitation by 2--3 times in these regions (Figs. \ref{fig:PrecipNAS}b and d). 

\begin{figure}
    \centering
    \includegraphics[width=\textwidth]{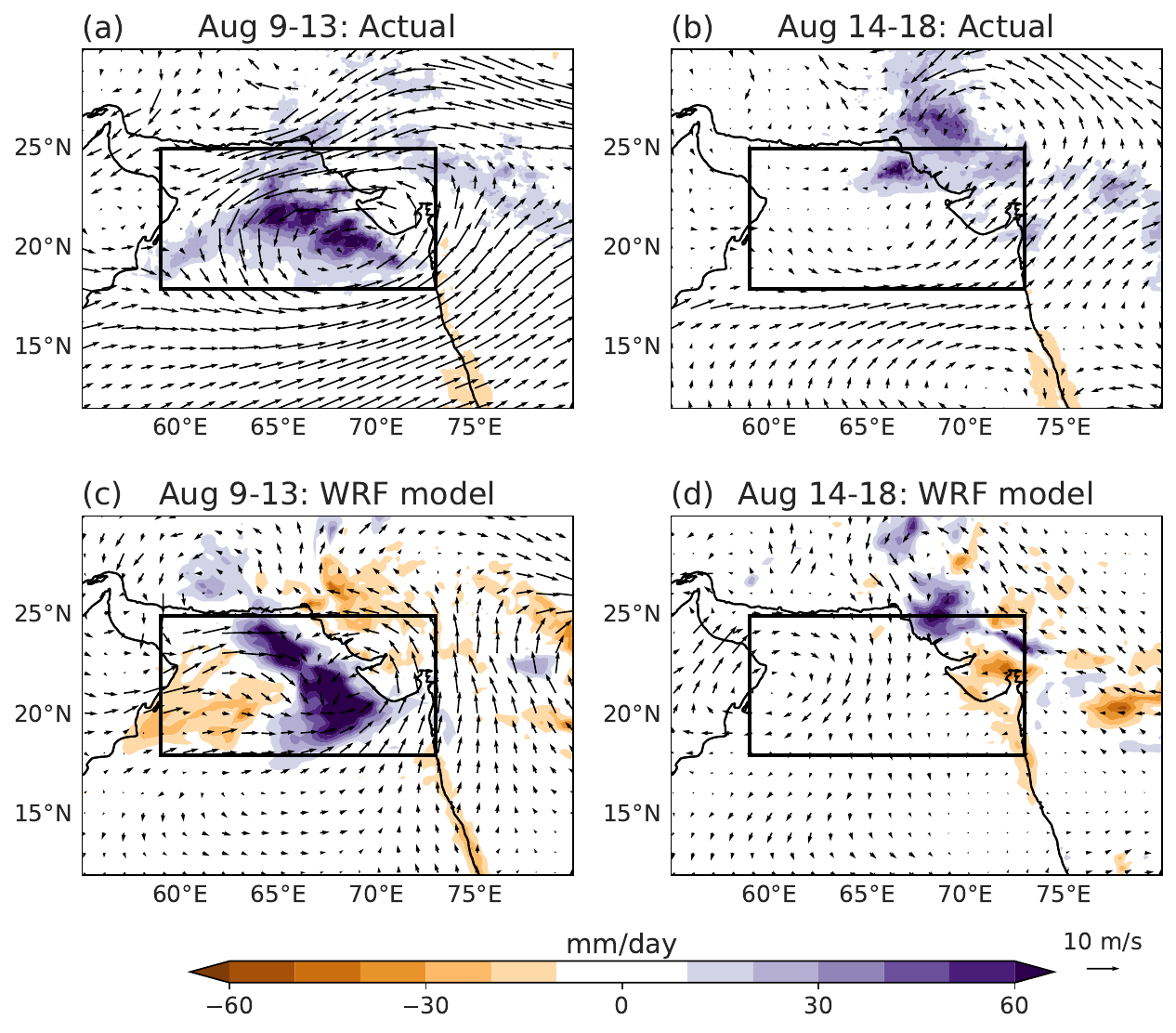}
    \caption{Anomalous precipitation (color) and 850 hPa winds (vectors) during the August 2022 floods in Pakistan. A MHW occurred in the northern Arabian Sea (black box), reaching its peak on August 8, 2022. The panels show the periods, (a) days 1 to 5, and (b) days 6 to 10, where day 0 marks the MHW peak. Precipitation is from IMERG and 850 hPa winds are from ERA5. Panels (c) and (d) show analogous plots from WRF model simulations (Main Run – Clim Run), with the WRF model initialized on August 4, 2022.}
    \label{fig:PrecipPak2022}
\end{figure}

The devastating floods in Pakistan in August 2022 appear to have been, at least in part, influenced by a MHW that occurred in the northern Arabian Sea. These catastrophic floods, caused by extreme precipitation events exceeding even the 99.9$^{\text{th}}$ percentile, affected approximately 33 million people in Pakistan and resulted in nearly 1500 fatalities \citep{nanditha2023pakistan}. Our analysis reveals that a MHW occurred in the northern Arabian Sea during this period, which reached its peak intensity on August 8, 2022. Over the next five days, a cyclonic circulation developed in the northern Arabian Sea, accompanied by strong precipitation (Fig. \ref{fig:PrecipPak2022}a). These anomalies continued to progress inland, triggering extreme precipitation events over Pakistan (Fig. \ref{fig:PrecipPak2022}b). Notably, the most intense 2-day precipitation occurred on August 17--18, 2022 \citep{nanditha2023pakistan}, coinciding with the presence of the low-level cyclone in the region (Fig. \ref{fig:PrecipPak2022}b). 

While the 2022 Pakistan floods have been linked to the triple-dip La Ni{\~n}a \citep{jeong2023triple}, atmospheric heatwaves over China and Europe \citep{hong2023causes}, and atmospheric rivers \citep{nanditha2023pakistan}, our findings emphasize the significant role of MHWs in the northern Arabian Sea in driving the extreme precipitation over Pakistan. The WRF model experiments further support this, showing that the intense low-level cyclone and inland precipitation occurs only when the warm SSTAs associated with the MHW is present (Figs. \ref{fig:PrecipPak2022}c and d), albeit with quantitative model-observation differences. Although seasonal prediction models have previously suggested that Arabian Sea SSTs influence forecasts of extreme precipitation over Pakistan \citep{doi2024seasonal}, this is the first study to explicitly emphasize the impact of MHWs. Recent research also indicates that the ongoing warming trend in the northern Arabian Sea, combined with a poleward shift of the low-level jet, strengthens cyclonic convergence and enhances intraseasonal precipitation trends over Pakistan \citep{li2022increase,li2023middle}. The similarity between the 2022 Pakistan floods and the composite response (compare Figs. \ref{fig:PrecipPak2022} and \ref{fig:PrecipNAS}) suggests that MHWs have the potential to trigger similar extreme precipitation events across northwestern India and Pakistan in a warming climate.

\section{Summary and Discussion}

Due to their profound socio-economic and ecological impacts, MHWs have been extensively studied in recent years, particularly in the context of anthropogenic warming \citep{capotondi2024global}. Located upstream of the Indian summer monsoon region, the Arabian Sea is a key moisture source for monsoon rainfall over the Indian subcontinent. Consequently, MHWs occurring during the boreal summer in the Arabian Sea not only affect marine ecosystems but can also influence summer monsoon precipitation. By analyzing satellite observations, ocean and atmosphere reanalysis datasets, high-resolution CMIP6 model simulations, and conducting numerical experiments using the WRF model, we identify the drivers of these MHW events and assess their impacts on regional weather patterns and monsoon rainfall over land.

To exclude the effect of warming trend (including both anthropogenic forcing and multi-decadal internal variability) on individual MHWs, we remove the SST trend---essentially linear over the study period (Supplementary Fig. S3b)---prior to identifying MHWs. Note that the typical MHW definition, based on SSTAs exceeding the 90$\mathrm{^{th}}$ percentile for at least five consecutive days \citep{hobday2016hierarchical}, imposes a lower bound of 5 days but no upper limit. However, our analyses show that long-lived and short-lived MHWs are triggered by different climatic phenomena operating at distinct timescales, making it necessary to examine them separately. We refer to long (short) MHWs as those lasting more (less) than 20 days. Varying this threshold by a few days yielded similar results. We find that most MHWs are short-lived: of the 32 (29) events detected in the eastern (northern) Arabian Sea during June–September of the study period, 25 (25) are short and 7 (4) are long MHWs. 

Individual MHWs are primarily driven by surface shortwave and latent heat fluxes (Fig. \ref{fig:SstBudget}). Horizontal advection and vertical mixing, however, also have some contributions. For long MHWs, their existence depends critically on interannual SSTA: none of the long MHWs exists without interannual SSTA, although intraseasonal SSTA induced by atmospheric intraseasonal variability (primarily BSISOs) can either weaken of enhance some events (Fig. \ref{fig:MhwCumIntYearly}). While the warm SSTAs induced by El Ni\~{n}o and positive IOD account for a large portion of the interannual SSTAs, other interannual climate variability independent of ENSO and IOD plays a comparable role (Figs. \ref{fig:MhwCumIntYearly}b-d). Sources of interannual climate variability other than ENSO and IOD remain to be explored. 

\begin{figure}
    \centering
    \includegraphics[width=\textwidth]{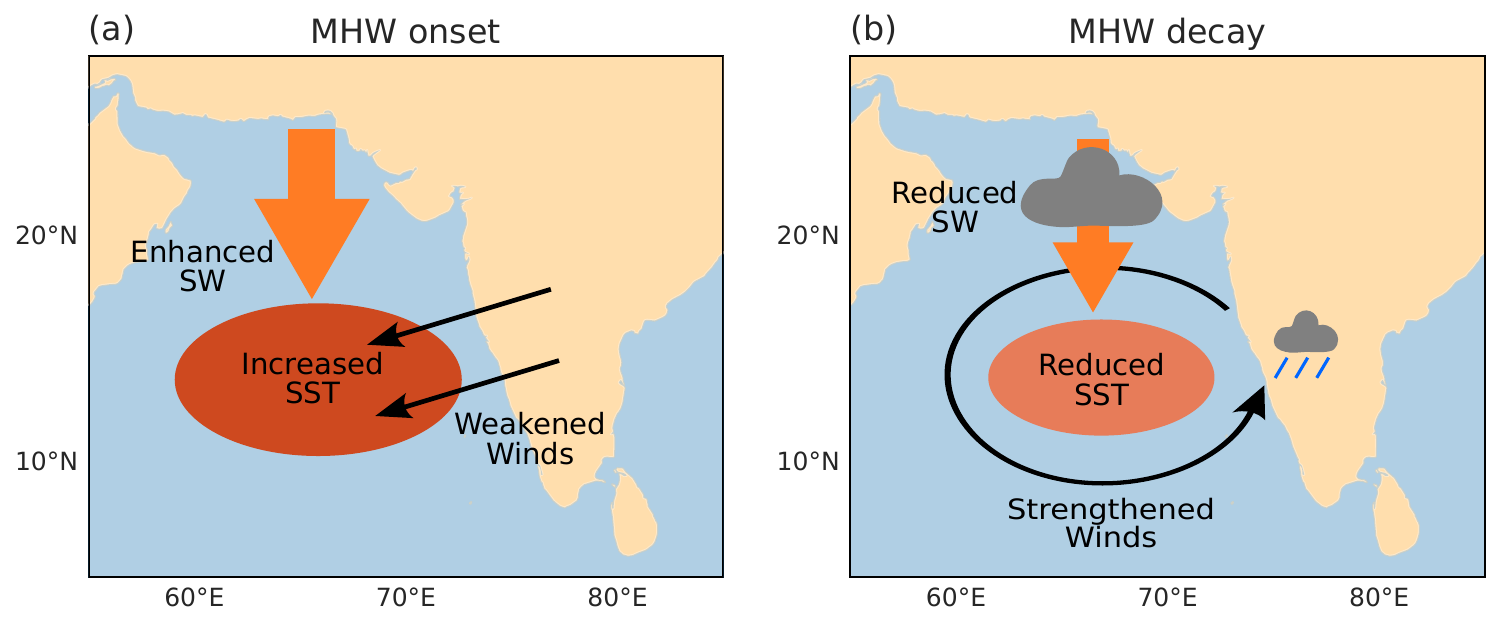}
    \caption{Schematic illustration of the drivers and impacts of MHWs during the (a) onset and (b) decay phases.}
    \label{fig:schematic}
\end{figure}

For short MHWs, intraseasonal SSTAs are crucial (Fig. \ref{fig:MhwCumIntYearly}e), with BSISOs acting as the primary driver. The underlying mechanism is schematically illustrated in Fig. \ref{fig:schematic}. These events are associated with the suppressed convection phases of BSISOs, characterized by clear skies and weakened winds, which enhance downward shortwave radiation and reduce evaporative heat loss, leading to SST warming (Figs. \ref{fig:schematic}a, \ref{fig:QnetEvoln}a, and \ref{fig:SstBudget}). Following this strong warming, a low-level cyclone develops, and the resulting convection and cloud cover reduce surface shortwave radiation, while strengthened winds on the equatorward side enhance evaporative heat loss. Together, these processes reduce SST, leading to the subsequent decay of the MHW (Figs. \ref{fig:schematic}b, \ref{fig:QnetEvoln}c and \ref{fig:SstBudget}). A similar process occurs in the northern Arabian Sea as well (compare Figs. \ref{fig:QnetEvoln} and S7, as well as the two regions in Fig. \ref{fig:SstBudget}).

While intraseasonal SSTAs are important for short MHWs, interannual SSTAs associated with ENSO, IOD, and other climatic phenomena also significantly modulate the year-to-year activities of short MHWs (Figs. \ref{fig:MhwCumIntYearly}b-d). Note that ENSO and IOD may also affect MHWs through modulating BSISOs, but this possibility is small since BSISO activities are similar during the positive and negative phases of ENSO and IOD (Fig. S4). BSISOs have two main spectral peaks: one at quasi-biweekly period (estimated by 10--25 day filtered fields) and the other at 30--60 day periods (estimated by 25--90 day filtered fields). Despite its shorter period, the quasi-biweekly mode is as important as the longer 30--60 day mode in generating intraseasonal SSTAs and triggering MHW events (Fig. S8).

MHWs, in turn, have significant impacts on regional weather and monsoon. The low-level cyclonic system that develops a few days after the MHW peak enhances convection and rainfall over the Arabian Sea and parts of the Indian subcontinent (Figs. \ref{fig:QnetEvoln}, \ref{fig:PrecipEAS}, \ref{fig:PrecipNAS} and \ref{fig:schematic}b). For MHWs that develop in the eastern Arabian Sea, the anomalous cyclone induces onshore (offshore) winds and moisture convergence (divergence) over the southern (northern) Western Ghats. This intensifies rainfall and increases the risk of extreme precipitation events along the southwest coast of India, while reducing it further north (Figs. \ref{fig:PrecipEAS} and \ref{fig:schematic}b). These cyclonic and precipitation anomalies are mainly associated with short-lived MHWs (Fig. S10). For MHWs of the northern Arabian Sea, they enhance precipitation and escalate the risk of extreme rainfall over the northern Arabian Sea, northwestern India, and Pakistan (Fig. \ref{fig:PrecipNAS}).  For instance, a MHW in the northern Arabian Sea during August 2022 induced a cyclonic circulation that subsequently propagated inland, triggering extreme precipitation and widespread flooding across Pakistan, affecting millions of people (Fig. \ref{fig:PrecipPak2022}). WRF model simulations for both the eastern and northern Arabian Sea show that such cyclonic circulation and rainfall anomalies develop only when MHWs are present, indicating that these are distinct responses driven by MHWs rather than an amplification of the response to BSISOs (Figs. \ref{fig:WrfPrecip} and \ref{fig:PrecipPak2022}c,d). That said, given the somewhat arbitrary nature of the MHW definition, relatively strong positive SST anomalies that do not meet the MHW threshold---but are still stronger than those associated with BSISO alone---may produce a qualitatively similar, albeit weaker, response. We have not explicitly tested the sensitivity of the atmospheric response to the amplitude of the SST forcing or the existence of any thresholds therein. Finally, we emphasize that the BSISO–MHW–precipitation mechanism described here applies only to short MHWs and does not represent long MHWs, which are more influenced by interannual SSTAs.

The impact of MHWs is further exacerbated by an apparent increase in the intensity of MHW-associated precipitation in recent years, particularly along the western coast of India and in northwestern India and Pakistan (Supplementary Fig. S11). The intensification of MHW-precipitation is much stronger than the overall upward trend of precipitation observed during the June--September monsoon season. Our analysis also reveals that the frequency of extreme precipitation events---the days exceeding the $95^{th}$ percentile---is increasing much more rapidly during MHWs compared to non-MHW days. However, given the relatively short span of the IMERG precipitation dataset (available only from 1998), it remains difficult to determine whether the recent increase in precipitation during MHWs (Figs. S8d–e) is driven by anthropogenic climate change, internal variability, or a combination of both. Furthermore, the intensification of BSISOs under warming SSTs is known to amplify intraseasonal precipitation \citep{sabeerali2014modulation, cheng2025increased}, which may have also contributed to the observed strengthening of MHW-associated rainfall in recent years. While long-term SST trends are removed here to isolate the drivers and impacts of MHWs \citep{smith2025baseline}, a comprehensive assessment of MHW risk under continued anthropogenic warming must also account for absolute SSTs and their spatial patterns, as these can modulate convective responses \citep{gadgil1984ocean, johnson2010changes, lindzen1987role, seo2023ocean}. Moreover, future changes in the dynamic–thermodynamic background state \citep{held2006robust, roxy2015drying}, combined with internal variability in mean-state conditions that affect Indian Ocean air–sea coupling, may further influence both the drivers and impacts of MHWs \citep{zhang2021decadal, zhang2024attributing}.

Given the crucial role of BSISOs in generating MHWs in the Arabian Sea and the demonstrated skill of BSISO forecasts \citep{fu2013intraseasonal, lee2015predictability}, it is feasible to develop a MHW forecast model to predict MHW events over the eastern and northern Arabian Sea. This, in turn, could help predict extreme precipitation events along the western coast of India, as well as in northwestern India and Pakistan. Such early warnings would facilitate timely mitigation efforts, helping to reduce the adverse impacts of MHWs on marine ecosystems and hazards associated with floods in the Indian summer monsoon region, thereby providing significant socio-economic and ecological benefits.

\acknowledgments

D.L.S. and W.H. are supported by NASA International Ocean Vector Wind Science Team award 80NSSC23K0982 and NASA Ocean Surface Topography Science Team award 0NSSC21K1190. T.S. is supported by NASA OVWST Grant 80NSSC23K0982, DOD Grant N00014-22-S-F008, and NSF Grant OCE 2242194. A.S. was supported by NASA (21-OSST21-0026) and DOE (DE-SC0024263). A.S. and R.S. also acknowledges the ONR ASTraL research initiative (N00014-23-1-2092). MB is supported by NASA Physical Oceanography support for the Ocean Vector Winds Science Team Leader (80NSSC24K0947).

\datastatement

OISST data used for identifying MHWs is available at \url{https://www.ncei.noaa.gov/products/optimum-interpolation-sst}. High-resolution CMIP6 model outputs were obtained from \url{https://pcmdi.llnl.gov/CMIP6/}. ERA5 data is available at  Copernicus Climate Change Service (C3S) Climate Data Store (CDS) website (\url{https://cds.climate.copernicus.eu/cdsapp#!/home}). CCMP surface wind data can be downloaded from \url{https://www.remss.com/measurements/ccmp/}. IMERG precipitation data is available at \url{https://gpm.nasa.gov/data/directory}. The Python scripts used for data analysis and the numerical model outputs presented in this study are available from the corresponding author upon request.

\bibliographystyle{ametsocV6}
\bibliography{refs.bib}

\end{document}